\documentclass[twocolumn,aps,prl,groupedaddress]{revtex4}
\usepackage{graphicx}
\usepackage{color}
\usepackage{amsmath}
\usepackage{enumitem}

\newcommand{\be}{\begin{equation}}
\newcommand{\ee}{\end{equation}}

\newcommand{\bea}{\begin{eqnarray}}
\newcommand{\eea}{\end{eqnarray}}

\newcommand{\p}{\partial}

\newcommand{\la}{\left\langle}
\newcommand{\ra}{\right\rangle}

\newcommand{\lp}{\left(}
\newcommand{\rp}{\right)}

\renewcommand{\vec}[1]{{\bf #1}}
\newcommand{\bra}[1]{\left\langle #1 \right|}
\newcommand{\ket}[1]{\left|#1\right\rangle}

\begin{document}
\title{Stokes Paradox, Back Reflections and Interaction-Enhanced Conduction}
%Carrier Collisions, Stokes Paradox and Higher-Than-Ballistic Conduction} % in Viscous Pinball Transport}
%Streaming Flows and Better-Than-Ballistic Conduction of Electron Fluids}
% Higher-Than-Ballistic Conductance of Electron Fluids}
%Super-ballistic transport in electron fluids}
%Viscous Point Contact} % and electron transport in graphene}

%\author{Haoyu Guo$^1$, Ekin Ilseven$^1$, Gregory Falkovich$^{2}$ and Leonid Levitov$^1$}
%
%\affiliation{$^1$Massachusetts Institute of Technology, Cambridge, Massachusetts 02139, USA \\ $^2$Weizmann Institute of Science, Rehovot 76100 Israel}

\author{Haoyu Guo$^1$, Ekin Ilseven$^1$, Gregory Falkovich$^{2,3}$ and Leonid Levitov$^1$}

\affiliation{$^1$Massachusetts Institute of Technology, Cambridge, Massachusetts 02139, USA \\ $^2$Weizmann Institute of Science, Rehovot 76100 Israel\\$^3$Institute for Information Transmission Problems, Moscow 127994 Russia}

%\date{\today}

\begin{abstract}
Interactions in electron systems can lead to viscous flows in which correlations allow electrons to avoid disorder scattering, reducing momentum loss and dissipation. We illustrate this behavior in a viscous pinball model, describing electrons moving in the presence of dilute  point-like defects.  Conductivity is found to obey an additive relation $\sigma=\sigma_0+\Delta\sigma$, with a non-interacting Drude contribution $\sigma_0$ and a 
contribution $\Delta\sigma>0$ describing conductivity enhancement due to interactions. 
The quantity $\Delta\sigma$ is enhanced by a logarithmically large factor originating from the Stokes paradox at the hydrodynamic lengthscales and, in addition, from an effect of repeated returns to the same scatterer due to backreflection in the carrier-carrier collisions occurring at the ballistic lengthscales. The interplay between these effects is essential at the ballistic-to-viscous crossover.
\end{abstract}
%Conductivity enhancement can be particularly dramatic in the hydrodynamic limit. 

%\pacs{}
\maketitle

Electron fluidity is a property of strongly interacting electron systems in which carrier movement resembles that of viscous fluids. 
Viscous electron flows are expected to occur in quantum-critical systems and in high-mobility conductors, so long as momentum-conserving electron-electron (ee) scattering dominates over 
other scattering processes\cite{gurzhi63,jaggi91,LifshitzPitaevsky_Kinetics,damle97}. 
Signatures of such flows have been
observed in 
ultra-clean GaAs, graphene and PdCoO${}_2$ \cite{dejong_molenkamp,bandurin2015,crossno2016,moll2016}. 
Electron fluids can exhibit a range of novel transport behaviors \cite{andreev2011,sheehy2007,fritz2008,muller2009,mendoza2011,forcella2014,tomadin2014,narozhny2015,principi2015,cortijo2015,lucas2016,LF,FL}.  
In particular, it has been predicted in the 1960's that viscosity can facilitate electron transport\cite{gurzhi63}. 
Furthermore, recently it was pointed out that electron fluid flowing through a constriction features conductance that exceeds the fundamental Landauer's ballistic bound\cite{guo2016}. Higher-than-ballistic conduction results from  correlations in a viscous flow that allow electrons to avoid scattering at the constriction boundary, thereby reducing dissipation due to momentum loss.

\begin{figure}
\includegraphics[scale=0.3]{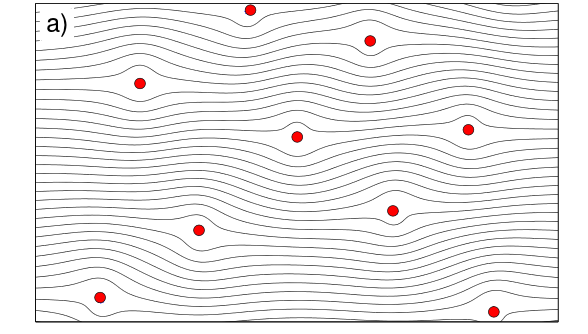}
\includegraphics[scale=0.25]{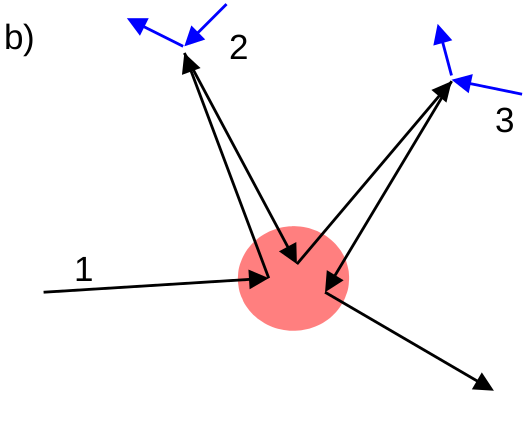} %{Fig1b_Stokes}
\caption{ 
Electron transport in the presence of point-like scatterers (red circles) facilitated by carrier-carrier collisions. 
a) Current streamlines for a viscous flow; 
the density of streamlines is proportional to the flow velocity. 
Streamlines bundle up, avoiding regions near the scatterers where momentum loss occurs. This enhances conductivity above the %ballistic-limit 
free-particle Drude value. b) Repeated returns of carrier 1 to the scatterer due to backreflection in the collisions with carriers 2 and 3. These processes, 
by preventing carriers 2 and 3 from reaching the scatterer, weaken the effective scatterer strength. % with regard to momentum loss.  
%Viscosity renormalizes the scatterer strength down, enhancing conductivity above the free-particle Drude value, Eq.\eqref{eq:sigma_log}.
}
\label{fig1}
\vspace{-5mm}
\end{figure}

The goal of this paper is to explore transport facilitated by electron viscosity in 
a two-dimensional system %``pinball'' dynamics 
in the presence of point-like scatterers, first introduced by Hruska and Spivak\cite{hruska02}  %\addQ{(various mathematical aspects of this regime have been further discussed in Refs.\cite{guo2016,lucas2016b}). }
%(see Fig.\ref{fig1}. 
In this case the reduction in resistance arises due to the collective behavior illustrated in Fig.\ref{fig1}a, wherein electron currents % bundle up to 
form streams that circumnavigate the regions near scatterers 
where momentum loss occurs.
This surprising behavior %, pointed out in Ref.\cite{hruska02}, 
is in a %stark 
departure from the common view that regards electron interactions as a hindrance to transport. 
%an impediment for transport.

As we will see, momentum-conserving ee collisions enhance conductivity by an {\it additive} viscosity-dependent contribution, $\sigma=\sigma_0+\Delta\sigma(\nu)$. 
Written explicitly, it is %equals
% giving
% predicting
%
\be\label{eq:sigma_log}
\sigma=\frac{N e^2 v^2}{2n_s } % U(\nu)}
\lp \frac1{U_0}+\frac1{4\pi \nu}\ln \frac{L}{a_*}\rp
%,\quad
%U(\nu)=\frac{U_0}{1+\frac{U_0}{4\pi \nu}\ln \frac{L}{a_*}}
%\mu=\frac{2ne}{n_s U(\eta)}
%,\quad U(\eta)=\frac{u_0}{1+\frac{u_0}{4\pi\eta}\ln \frac{\xi}{a}}
,
\ee
%
% which describes the proportionality between the applied electric field and the drift velocity.
where $U_0$ is the bare scatterer strength, $n_s$  is the scatterer concentration, $N$ is the density of states, $v$ is Fermi velocity, $\nu$ is viscosity, see Eq.\eqref{eq:viscous_mode}. The two terms in Eq.\eqref{eq:sigma_log} represent the free-particle Drude contribution $\sigma_0$ and the viscous contribution $\Delta\sigma$, respectively. 
The contribution $\Delta\sigma$ is enhanced by a log factor in which $L$ is the system size or the distance between scatterers $n_s^{-1/2}$, whichever is the smallest, and 
$a_*=(a\ell_{\rm ee})^{1/2}$, %where 
with $a$ the scatterer radius and $\ell_{\rm ee}$ the ee collision mean free path. 
The lengthscale $a_*$ is such that the diffusion time $a_*^2/\nu=4a_*^2/v\ell_{\rm ee}$ is comparable to the  ballistic time $a/v$. 

The log enhancement arises as a combination of two distinct effects. One is a log factor $\ln(L/\ell_{\rm ee})$ originating at the hydrodynamic lengthscales $r\gtrsim\ell_{\rm ee}$ %in the long-wavelength limit %the log factor originates from the Stokes logarithm familiar 
from the two-dimensional Stokes paradox, as pointed out in Ref.\cite{hruska02}.  
 %by Hruska and Spivak\cite{hruska02}. 
The other is a log %peculiar logarithmic 
factor of the form $\frac12\ln(\ell_{\rm ee}/a)$, originating from repeated scattering processes illustrated in Fig.\ref{fig1}b. These processes describe multiple returns of one carrier to the same scatterer due to backreflection in the ee collisions occurring at the ballistic lengthscales $a\lesssim r\lesssim \ell_{\rm ee}$, see Eq.\eqref{eq:return_prob}. Naturally, such % multiple return 
%of one carrier to the same scatterer
processes translate into a reduction in the probability for other carriers to reach the scatterer. 
% As we will see, %discussed below, 
The two log contributions 
combine additively to generate the term $\ln(L/a_*)$ % contribution 
in Eq.\eqref{eq:sigma_log}. 

The above result, %described by 
Eq.\eqref{eq:sigma_log}, holds for the ee collision rate in a wide range, $a\lesssim \ell_{\rm ee}\lesssim L$, mapping out the viscous-to-ballistic crossover. Notably, because of the ballistic contribution to the log factor, % arising from the ballistic lengthscales, 
the effect of conductivity enhancement by interactions %remains prominent 
survives even for the ee mean free path values as large as $L$. In this case the Stokes log factor vanishes, however the backreflection log factor reaches maximum [with $a_*\approx (aL)^{1/2}$]. 
%We analyze the problem using 
In our analysis we will use the method of {\it quasi-hydrodynamic variables}\cite{levinson77} which employs the moments of particle distribution function conserved in ee collisions. 
%This method will allow us to explore the crossover from hydrodynamic to free-particle transport. 
% In our analysis we 
We will ignore the effect of momentum relaxation due to electron-phonon scattering. This effect can be easily modeled by adding a damping term to the transport equations, e.g. see Refs.\cite{LF, bandurin2015,lucas2016b}. Also, we assume that dephasing due to finite temperature suppresses the interference effects at distances greater than $a$, which allows us to use the incoherent transport picture at such lengthscales. 

The effect of viscosity in Eq.\eqref{eq:sigma_log} can be understood as a renormalization of the scatterer strength %described by
% change of the bare scatterer strength $U_0$ to an effective strength
\be\label{eq:U(nu)}
U(\nu)=\frac{U_0}{1+\frac{U_0}{4\pi \nu}\ln \frac{L}{a_*}}
.
\ee
% Physically, renormalization 
In a hydrodynamic picture, suppression of $U$ originates from currents streaming to avoid scatterers 
and creating stagnation regions near the scatterers that act as a `lubricant' to diminish momentum loss and facilitate transport. The backreflection processes (Fig.\ref{fig1}b) contribute in a similar way albeit at the ballistic lengthscales. 
The effective scatterer strength becomes weaker 
as the system becomes more fluid i.e. when the mean free path $\ell_{\rm ee}$ (and thus the value of $\nu$) %=\frac14 v_F\ell_{\rm ee}$)
decreases. 

The log divergence in Eq.\eqref{eq:sigma_log} arises from 2D momentum diffusion in a manner reminiscent of the seminal log divergences due to 2D diffusion in quantum-coherent transport  (weak localization and related effects \cite{abrahams79,gorkov79,altshuler79}). However, here we find a logarithmic enhancement rather than a suppression of conductivity. Also, while the log divergences dominate in quantum-coherent transport at low temperature, here it becomes more prominent as the system becomes more fluid with temperature growing. Eq.\eqref{eq:sigma_log} also indicates that the log divergence `amplifies' the viscosity dependence, which becomes prominent once $\ell_{\rm ee} < a\ln(L/a_*)$ 
(see below).

% In developing a model of transport in the presence of
Since transport in our system is dominated by momentum-conserving collisions it is convenient to work with the quasi-hydrodynamic variables defined as the deviation in the average 
particle density and momentum from local equilibrium \cite{levinson77}. 
%In our analysis 
Here we will use Boltzmann kinetic equation linearized in deviations of particle distribution from the %equilibrium 
Fermi step (assuming $k_{\rm B}T\ll E_F$),
\be\label{eq:Boltzmann}
\lp\p_t+\vec v\nabla_{\vec x}\rp f(\theta,\vec x,t)=I_{\rm ee}(f)+I_{\rm dis}(f)
,
\ee
were $\theta$ is the angle parameterizing momentum at the 2D Fermi surface. The collision operators $I_{\rm ee}$ and $I_{\rm dis}$ describe the carrier-carrier and disorder scattering. Particle collisions conserve the particle number and momentum, which provide quasi-hydrodynamic variables for our problem. We express these quantities, which are the zero modes of $I_{\rm ee}$, as angular harmonics of % momentum distribution. 
the distribution $f(\theta,\vec x,t)$:
\be\label{eq:zero_modes}
f_0=\la f(\theta)\ra_\theta
,\quad
f_{\pm 1}=\la e^{\mp i\theta}f(\theta)\ra_\theta
\ee
where we introduced notation $\la...\ra_\theta= \oint ...\frac{d\theta}{2\pi}$. Disorder collisions, in contrast, conserve $f_0$ but not $f_{\pm 1}$.
To facilitate the analysis, we choose a model for $I_{\rm ee}$ and $I_{\rm dis}$
with a single relaxation rate for all non-conserved harmonics:
\be\label{eq:Iee_Idis}
I_{\rm ee}(f)=-\gamma(f-Pf)
,\quad
I_{\rm dis}(f)=-\alpha(\vec x)P'f
,
\ee
where $\gamma$ represents the ee collision rate, and
\be
\alpha(\vec x)=\sum_s u(\vec x-\vec x_s)
\ee 
describes randomly-placed  scatterers. Here $P=\ket{0}\bra{0}+\ket{1}\bra{1}+\ket{-1}\bra{-1}$  is the projector on the angular harmonics \eqref{eq:zero_modes}, whereas $P'$, defined in a similar manner,  projects on the $m=\pm1$ harmonics. The quantity $Pf$ in $I_{\rm ee}$ then stands for
\be
Pf(\theta)={\sum}_{m=0,\pm1}
\la e^{im(\theta-\theta')}f(\theta')\ra_{\theta'}
\ee
and $P'f$ is given by a similar expression with $m=\pm1$. The form of $I_{\rm ee}$ and $I_{\rm dis}$ in Eq.\eqref{eq:Iee_Idis} ensures momentum conservation in particle collisions and %the lack of it 
momentum loss in the disorder collisions.

We start with analyzing the hydrodynamic modes of Eq.\eqref{eq:Boltzmann} in the absence of disorder, $I_{\rm dis}=0$.
In this case, Eq.\eqref{eq:Boltzmann} takes the form $(\hat K-\gamma P)f=0$, where $\hat K=\p_t+\vec v\nabla_{\vec x}+\gamma \hat 1$.
Since $f_0$ and $f_{\pm1}$ are zero modes of the particle collision operator  $I_{\rm ee}$, they dominate at low frequencies and long wavelengths. Accordingly, we can obtain hydrodynamic modes
from plane-wave solutions, $f(\theta,\vec x,t)\sim f(\theta) e^{-i\omega t+i\vec k\vec x}$. Solving  Eq.\eqref{eq:Boltzmann} as $f=\gamma \hat K^{-1}Pf$ we project $f$ on  the harmonics $f_0$ and $f_{\pm1}$. This gives three coupled equations $f_m=g_{mm'}f_{m'}$, where $g_{mm'}=\la m|\gamma P\hat K^{-1}P|m'\ra$. Direct calculation gives
\be\label{eq:g3x3}
g_{mm'}=\la \frac{\gamma e^{i(m-m')\theta}}{\gamma_\omega+i\vec k\vec v}\ra_\theta
\!\!
=\tanh\beta \frac{\gamma e^{i\theta_k\Delta m} }{\gamma_\omega\lp ie^{\beta}\rp^{|\Delta m|}}
%\frac{\gamma e^{i\theta_k(m-m')} }{\gamma_\omega\lp ie^{\beta}\rp^{|m-m'|}}
.
\ee
Here $\gamma_\omega=\gamma-i\omega$, $\sinh\beta=\frac{\gamma_\omega}{kv}$ and $\Delta m=m-m'$, $m,m'=0,\pm 1$. The quantity in Eq.\eqref{eq:g3x3} is evaluated by writing $\vec k\vec v=kv\cos\tilde\theta$, with $\tilde\theta=\theta-\theta_k$ the angle between particle velocity $\vec v$ and momentum $\vec k$, and integrating over $\theta$. As we now show, the equations
$f_m=g_{mm'}f_{m'}$ generate an acoustic and a viscous mode.

The $3\times 3$ matrix $g_{mm'}$ can be brought to a block-diagonal form by taking into account that the acoustic mode is longitudinal whereas the viscous mode is transverse with respect to $\vec k$. Transforming to the even/odd basis 
\be
\ket{0},\quad \ket{c}=\frac{\ket{1_k}+\ket{-1_k}}{\sqrt{2}}
,\quad
\ket{s}=\frac{\ket{1_k}-\ket{-1_k}}{\sqrt{2}}
,
\ee
where $\ket{m_k}=e^{-im\theta_k}\ket{m}$. 
The even and odd modes
%$\left.|c\ra$ and $\left.|s\ra$
correspond to $f_c(\theta)\sim\cos \tilde\theta$ and $f_s(\theta)\sim\sin \tilde\theta$.
The odd-mode $1\times 1$ block gives $g_{ss}=\frac{\gamma}{\gamma_\omega}\tanh\beta (1+e^{-2\beta})$. Taylor-expanding the dispersion relation $1=g_{ss}$ in small $\omega$ and $k$ yields the viscous mode %dispersing as
\be\label{eq:viscous_mode}
\omega=-i\nu k^2
,\quad
\nu=v^2/4\gamma
.
\ee
Here $\nu$ is the viscosity defined so that the dispersion in Eq.\eqref{eq:viscous_mode} agrees with that obtained from linearized Navier-Stokes equation $(\p_t-\nu\nabla^2)\vec v=-\nabla P$.

The acoustic mode can be obtained from the even-mode $2\times2$ block
% The even-mode $2\times2$ block is %we find
\be
\lp\begin{array}{cc} g_{00} & g_{0c} \\ g_{c0} & g_{cc}\end{array}\rp
=
\frac{\gamma\tanh\beta}{\gamma_\omega}
\lp\begin{array}{cc} 1 & -i\sqrt{2} e^{-\beta} \\  -i\sqrt{2} e^{-\beta} & 1-e^{-2\beta}\end{array}\rp
.
\ee
The dispersion relation ${\rm det}\,(1-g)=0$, Taylor-expanded in  $\omega$ and $k$, yields a damped acoustic mode  
% $\omega=\frac1{\sqrt{2}}kv-\frac{i}2\nu k^2$.
$\omega=kv/\sqrt{2}-i\nu k^2/2$.
% ( "with viscosity $\nu=v^2/4\gamma$" moved to front.)

Next we analyze the effect of disorder $\alpha(\vec x)$.
The state describing uniform current $j$ in the absence of disorder is $f^{(0)}=2j\cos\theta$ (without loss of generality we consider current flowing in the $x$ direction). The distribution perturbed by disorder
satisfies Eq.\eqref{eq:Boltzmann} which we write in the Fourier representation setting $\omega=0$ for a steady state
\be\label{eq:G(f0+df)}
\lp % -i\omega+
i\vec k\vec v+\gamma \hat 1 -\gamma P+\hat\alpha P'\rp \ket{ f^{(0)}+\delta f }=0
.
\ee
% where $\tilde\theta=\theta-\theta_k$ is the angle between particle velocity and momentum $\vec k$.
Here we treat the disorder scattering term as an operator in momentum representation, 
\be
\la \vec k|\hat\alpha|\vec k'\ra=\sum_s e^{i\vec x_s(\vec k-\vec k')} u_{\vec k-\vec k'},
\ee   
where $u_{\vec k}$ is the Fourier transform of $u(\vec x)$. Taking into account that $f_0$
%is a steady state of the equation
satisfies Eq.\eqref{eq:G(f0+df)} in the absence of disorder, $\alpha=0$, we write a formal solution of Eq.\eqref{eq:G(f0+df)}
as
\be\label{eq:df}
\ket{\delta f}=-G\lp \hat\alpha -\hat\alpha G\hat\alpha +\hat\alpha G\hat\alpha G\hat\alpha -... \rp \ket{f^{(0)}}
.
\ee
Here $G=1/(\gamma+i\vec k\vec v-\gamma P)$ and, for conciseness, we absorbed the projector $P'$ into $\hat\alpha$.

% Quasi-hydrodynamic variables are obtained by 
 % It will be convenient to 
Next we project $\ket{\delta f}$ on the quasi-hydrodynamic subspace of $m=0,\pm1$ harmonics. Acting on $\left. |\delta f\ra$ in Eq.\eqref{eq:df} with $P$ we can write the result as
\be\label{eq:df}
P \ket{ \delta f}= -D\lp \hat\alpha-\hat\alpha D\hat\alpha +\hat\alpha D\hat\alpha D\hat\alpha -... \rp \ket{f^{(0)}}
\ee
where $D=PGP$ is a $3\times3$ matrix in the $m=0,\pm1$ space (here we used the identity $\hat\alpha=P\hat\alpha P$ which follows from $PP'=P'P=P'$).

The matrix $D$ can be expressed through the matrix $g$ given in Eq.\eqref{eq:g3x3} by setting $G_0=1/(i\vec k\vec v+\gamma)$ and performing an expansion of $G=1/(G_0^{-1}-\gamma P)$ in $\gamma P$: % to obtain
\be
G=\hat K^{-1}=G_0+G_0TG_0
,\quad
T=\frac{\gamma P}{1-\gamma PG_0P}
.
\ee
Here we resummed the series, expressing the result in terms of a $3\times 3$ matrix $T$ in a manner analogous to the derivation of the Lippmann-Schwinger $T$-matrix for quantum scattering with a finite number of active channels.
We note that $\gamma PG_0P$ is nothing but the matrix $g$ in Eq.\eqref{eq:g3x3} taken at $\omega=0$. Plugging this into $D=PGP$ and performing matrix inversion we obtain a relation
\be %\nonumber
D=\frac{\gamma^{-1}g}{1-g}
=\frac{\sinh\beta}{\gamma} \!\lp\begin{array}{ccc} e^{\beta} & -i z_k &-e^{\beta} z_k^2\\ -i\bar z_k & e^{-\beta} & -i z_k \\ -e^{\beta} \bar z_k^2 & -i \bar z_k & e^{\beta} \end{array}\rp
%=\frac{\sinh\beta}{\gamma} \lp\begin{array}{ccc} e^{\beta} & -i e^{i\theta_k} &-e^{\beta} e^{2i\theta_k}\\ -ie^{-i\theta_k} & e^{-\beta} & -ie^{i\theta_k} \\ -e^{\beta}e^{-2i\theta_k} & -ie^{-i\theta_k} & e^{\beta} \end{array}\rp
%\!,\ \ z_k=e^{i\theta_k}
,
\ee
where $z_k=e^{i\theta_k}$.

Plugging $D$ in Eq.\eqref{eq:df} and reinstating $P'$ in $\hat\alpha$ we evaluate the quantity $\hat\alpha D\hat\alpha$ for a single point-like 
scatterer. Writing $\alpha(\vec x)=u(\vec x)P'$ and taking 
into account that $P'DP'$ eliminates the middle row and column in $D$,
%and using a short-range scatterer  Fourier transform $u_0=u_{\vec k= 0}$, gives
we obtain
\be\label{eq:aDa}
\hat\alpha D\hat\alpha=u_0^2 \int %(d^2k) 
\frac{d^2k}{(2\pi)^2}
\frac{\sinh\beta}{\gamma}  \lp\begin{array}{cc} e^{\beta} &-e^{\beta} z_k^2\\ -e^{\beta}\bar z_k^2 & e^{\beta} \end{array}\rp
.
\ee
Here, anticipating that the contribution $\hat\alpha D\hat\alpha$ is dominated by lengthscales $r\gg a$, %$(d^2k)$ stands for $d^2k/(2\pi)^2$ and 
we approximated the Fourier transform of a scatterer as $u_0=u_{\vec k= 0}$. Next we note that, due to azimuthal symmetry, the integral of the terms $z_k^2$ and $\bar z_k^2$ vanishes. The integral in Eq.\eqref{eq:aDa} then yields the projector $P'$:
\be
\hat\alpha D\hat\alpha=P' u_0^2 I(\nu),\quad
I(\nu)=\int_{1/L}^{1/a} \frac{d^2k}{(2\pi)^2}
\frac{\sinh\beta e^{\beta}}{\gamma}
,
\ee
where we expressed the UV and IR cutoffs through the scatterer radius $a$ and the distance between the scatterers $L$, respectively. 
% $L$ is the system size or the distance between scatterers, whichever is smaller. 

Further, after the replacement $u_0=u_{\vec k= 0}$, all higher-order terms in the series \eqref{eq:df} can be evaluated in a similar manner since momentum integration in each of the $D$ blocks can be performed independently. Summing the series gives an effective scatterer strength renormalized by viscosity
\be\label{eq:U(nu)_I(nu)}
U(\nu)=\frac{u_0}{1+u_0 I(\nu)}
.
\ee
The integral $I(\nu)$ can be evaluated exactly, giving
\be\label{intk_exact}
\begin{split}
&I(\nu)=\int \frac{d^2k}{(2\pi)^2}\frac{\sinh\beta e^{\beta}}{\gamma} =\frac1{2\pi v\ell_{\rm ee}}\lp\sqrt{1+k^2\ell_{\rm ee}^2} %\frac{k^2 v^2}{\gamma^2}}
\right. \\ &\left. +2\ln\lp k\rp-\ln\lp 1+\sqrt{1+k^2\ell_{\rm ee}^2} %\frac{k^2 v^2}{\gamma^2}}
\rp \rp\Big|_{1/L}^{1/a}
,
\end{split}
\ee
where $\ell_{\rm ee}=v/\gamma$. 

Eq.\eqref{eq:U(nu)_I(nu)} describes several regimes of interest. First we consider the {\it hydrodynamic} regime when the ee collision mean free path is much smaller than the scatterer radius, $\ell_{\rm ee}\ll a \ll L$. In this case %we have 
\be\label{eq:I1}
I(\nu)\approx \frac1{4\pi\nu}\ln\frac{L}{a}
.
\ee 
% and 
% In the hydrodynamic regime 
Eq.\eqref{eq:U(nu)_I(nu)} then gives a renormalized scatterer strength %equals
% hydrodynamic regime we find
%$1/L\ll 1/a\ll \gamma/v$, consequently,
\be\label{eq:U(nu)u0}
U(\nu)=\frac{u_0}{1+\frac{u_0}{4\pi \nu} \ln \frac{L}{a}}
,
\ee
%Note that this result agrees with Eq.\eqref{eq:U(eta)} which is derived from Navier-Stokes equation.
%
which is nothing but the renormalization by the Stokes logarithm 
% a result that agrees with that 
derived in Ref.\cite{hruska02}. Next we consider the {\it ballistic} free-particle regime, $\ell_{\rm ee}\gg L\gg a$. Taking the limit $\gamma\to 0$, $\ell_{\rm ee}\to\infty$, and expanding Eq.\eqref{intk_exact} to leading order in $1/a$ and $1/L$, we find 
\be\label{eq:I2}
I(\nu)=\frac1{2\pi v a}
.
\ee
Plugged in Eq.\eqref{eq:U(nu)_I(nu)} it gives a $\nu$-independent result
\be\label{eq:U0}
U_0=\frac{u_0}{1+\frac{u_0}{2\pi v a}}
.
\ee
%where $k_>=a^{-1}-L^{-1}\approx a^{-1}$ is the UV cutoff. 
We note that, strictly speaking, in this limit there is no small parameter allowing us to perform %Here we performed 
summation of perturbation series \eqref{eq:df} by treating $u_{\vec k-\vec k'}$ as momentum-independent and decoupling different momentum integrals. However, %one can argue that, 
while a more careful approach may % Relaxing this assumption may 
generate a numerical prefactor before $\frac{u_0}{2\pi v a}$, this will not affect the resulting general behavior of $U_0$.

Next, we use the above results to analyze the viscous-to-ballistic crossover regime $a\ll \ell_{\rm ee} \ll L$. In this case % we have
\be
I(\nu)\approx \frac1{2\pi v a}+\frac1{2\pi v\ell_{\rm ee}}\lp 2\ln\frac{L}{a} -\ln\frac{\ell_{\rm ee}}{2a} -1\rp
.
\ee
It is instructive to separate the ballistic and the hydrodynamic contributions, found above, and write $I(\nu)$ as % this result as
\be\label{eq:I3}
I(\nu)\approx I_1+I_2+ 
%\frac1{2\pi v a}+\frac1{\pi v\ell_{\rm ee}}\ln\frac{L}{\ell_{\rm ee}}+
\frac1{2\pi v\ell_{\rm ee}} \ln\frac{2\ell_{\rm ee}}{ea} 
\quad (e=2.71828...)
,
\ee
where %$e=2.71828...$ and  
$I_1$, $I_2$ are given by Eqs.\eqref{eq:I1},\eqref{eq:I2}. Besides the ballistic and  hydrodynamic contributions $I_{1,2}$, the function % which are identical to those found above, 
$I(\nu)$ contains a new term $\frac1{2\pi v\ell_{\rm ee}}
\ln\frac{2\ell_{\rm ee}}{ea}$ which, as we will see, alters the behavior in a interesting way.

The meaning of this term %in Eq.\eqref{eq:I3} 
can be understood by considering Eq.\eqref{eq:U(nu)_I(nu)} which describes an effective scatterer strength renormalized by repeated return processes. The contributions $I_1$ and $I_2$ describe returns from the `inner' lengthscales $r\sim a$ and from the hydrodynamic lengthscales $\ell_{\rm ee}\lesssim r\lesssim L$. The last term in Eq.\eqref{eq:I3} therefore describes returns from the lengthscales $a\lesssim r\lesssim \ell_{\rm ee}$. In this vein, the origin of the log factor $\ln\frac{\ell_{\rm ee}}{a}$ is explained by a simple physical argument. Consider a carrier that, after first scattering event, travels away from the scatterer and, % Suppose this carrier, 
at a distance $r$, collides with another carrier and bounces back to the scatterer, as illustrated in Fig.\ref{fig1}b. The probability for such a process is estimated as
\be\label{eq:return_prob}
p\sim\frac{\gamma}{v}\int_a^{\ell_{\rm ee}}dr  \frac{\Delta\theta(r)}{2\pi}
\ee
where $\Delta\theta(r)\sim \frac{a}{r}$ is the angle at which the scatterer is seen from a distance $r$. 
%Up t o an order-one factor under the log, 
Integration over $r$ gives a log factor identical to that in Eq.\eqref{eq:I3}. 

The impact of the term  $\frac1{2\pi v\ell_{\rm ee}}
\ln\frac{2\ell_{\rm ee}}{ea}$ on scatterer renormalization can be clarified by rewriting  Eq.\eqref{eq:I3} as
\be
I(\nu)=
\frac1{2\pi v a} + \frac1{4\pi\nu} \ln \frac{L}{a_*}
,\quad
a_*=\sqrt{\frac{e}2 a\ell_{\rm ee}}
%a_*=\sqrt{\frac{e}2 a\ell_{\rm ee}}
,
\ee
i.e. the UV cutoff lengthscale shifts to the value $a_*$ much smaller than the hydrodynamic cutoff lengthscale $\ell_{\rm ee}$. Since $a\ll a_*\ll \ell_{\rm ee}$, the effect of the cutoff $a_*$ is particularly important in the viscous-to-ballistic crossover regime. Namely, even $\ell_{\rm ee}$ approaches $L$ the value $a_*$ continues to be small compared to $L$ and therefore the log $\ln\frac{L}{a_*}$ continues to be large. 
%that in the Stokes contribution, Eq.\eqref{eq:U(nu)u0}. %$a\ll a_*\ll \ell_{\rm ee}$
%in Eq.\eqref{eq:I3} can be understood by rewriting 

To distill the dependence on viscosity we consider the renormalized scatterer strength, Eq.\eqref{eq:U(nu)_I(nu)}.
Plugging $I(\nu)$ above and expressing $\ell_{\rm ee}$ through viscosity \eqref{eq:viscous_mode},
%value $\nu=v^2/4\gamma$ obtained above,
gives %that the scatterer strength  is {\it renormalized down}: % as
\be\label{eq:U(nu)_new}
U(\nu)=\frac{u_0}{1+\frac{u_0}{4\pi \nu} \ln \frac{L}{a_*}
+\frac{u_0}{2\pi v a}}
=\frac{U_0}{1+\frac{U_0}{4\pi \nu} \ln \frac{L}{a_*}}
.
\ee
Here, to clarify the dependence on $\nu$, we expressed $u_0$ through the renormalized scatterer strength $U_0$ found in the ballistic regime, Eq.\eqref{eq:U0}. This gives to the dependence in Eq.\eqref{eq:U(nu)}. As a function of viscosity, $U(\nu)$ varies from $U_0$ in the ballistic limit ($\ell_{\rm ee}\approx L$) down to $4\pi\nu/\ln (L/a_*)$ in the highly viscous limit ($\ell_{\rm ee}\approx a$). Similar behavior is found as a function of $L$: from $ U_0$ for $L\sim a$ down to  $ 4\pi\nu/\ln (L/a_*)$ at very large $L$.

Lastly, we use the above results
%evaluate density perturbation around the scatterer and use it
to relate the carrier drift velocity and the electric field. From Eq.\eqref{eq:df} the density perturbation around one scatterer is given by
\be
f_0(k)=\frac{U(\nu)}{iv}\lp \frac{f_1}{k_1-ik_2} +{\rm c.c.}\rp
=\frac{2jk_1U(\nu)}{iv(k_1^2+ k_2^2)}
.
\ee
To restore physical units we scale $j$ by $ev$. The current-induced potential for one scatterer is found  by dividing $f_0$ by the density of states $N=dn/d\mu$.
Evaluating the electric field as $\vec E(k)=-i\vec k f_0(k)/N$, we perform spatial averaging (by setting $k_2=0$) and multiply by the density of scatterers $n_s$. This gives the current-field relation $eE=\frac{2n_s U(\nu)}{N v^2}j$. Expressing $j$ through the drift velocity as $en v_{\rm d}$ we obtain the conductivity given in Eq.\eqref{eq:sigma_log}.

The log-divergent suppression of scattering and associated enhancement of conductivity can be linked to the hydrodynamic modes discussed above. In particular, as shown in Appendix, 
%Supporting Information, 
%the log divergence and 
the dependence on the IR cutoff $L$ in Eqs.\eqref{eq:sigma_log},\eqref{eq:U(nu)} can be fully understood in terms of the viscous mode alone, and reproduced using the Navier-Stokes hydrodynamics. This approach, however, fails to generate the correct UV cutoff dependence on the e-e scattering rate, $a_*\sim (\ell_{\rm ee} a)^{1/2}$. The latter originates from repeated scattering processes induced by backreflection in the ee collisions. The corresponding lengthscales are ballistic rather than hydrodynamic. 
%. The unusual square-root scaling indicates 
%% dependence indicates 
%that the acoustic mode plays an important role at the UV scale.

The log divergence amplifies the viscosity-dependent enhancement of conductivity, which becomes prominent for $\nu$ smaller than $(U_0/4\pi)\ln(L/a_*)$. Since $U_0\approx 2\pi v a$ for a strong scatterer, where $a$ is the scatterer size, conductivity increases well above the non-interacting value once $\ell_{\rm ee}< a\ln(L/a_*)$ i.e. already in the weakly-interacting regime. This behavior facilitates reaching the viscous regime and probing the viscous-to-ballistic crossover.

\section{Acknowledgements}
We acknowledge support of the Center for Integrated Quantum Materials (CIQM) under NSF award 1231319
(L.L.), partial support by the U.S. Army Research Laboratory and the U.S. Army Research Office
through the Institute for Soldier Nanotechnologies, under contract number W911NF-13-D-0001 (L.L.), 
MISTI MIT-Israel Seed Fund (L.L. and G.F.), 
%This work is supported by
the  Israeli Science Foundation (grant 882) (G.F.) and the Russian Science Foundation (project 14-22-00259) (G.F.).

%\end{document}

%\newpage

%\section{Supporting Information for ``Higher-Than-Ballistic Conduction of Viscous Electron Flows'' by Haoyu Guo, Ekin Ilseven, Gregory Falkovich and Leonid Levitov}

\section{ Appendix: 
The hydrodynamic %origin of the 
log divergence}

%\addHG{In fact, the IR part log renormalization factors Eq.\eqref{eq:U(nu)U0} and Eq.\eqref{eq:U(nu)u0} arise from the hydrodynamical modes described before. In what follows, we will derive the IR renormalization through the simpler linearized Navier-Stokes type equation}
%Transport in the pinball model (Fig.\ref{fig1}b) can be modeled in a simple  way by a linearized Navier-Stokes-type equation
Here we consider transport in the pinball model using a hydrodynamic approach. Granted, such an approach treats the electron mean free path as the shortest scale in the problem, and thus is inadequate to describe the viscous-to-ballistic crossover. However, hydrodynamics provides a simple interpretation
%as we will see, it helps to clarify the origin
of the logarithmic renormalization of the scatterer strength found from the kinetic equation approach. Below we show that this renormalization can be accounted for by the viscous modes described by a linearized Navier-Stokes equation
\be
\eta\nabla^2 v_i =\alpha(\vec r) v_i+ ne\p_i \phi
,\quad
\alpha(\vec r)=\sum_s u(\vec r-\vec r_s)
\label{S1}
\ee
 where $\phi$ is electric potential and the terms $u(\vec r-\vec r_s)$ describe randomly placed  scatterers. The flow velocity obeys the incompressibility condition ${\rm div}\,\vec v=0$ and is related with electric current through $\vec j=ne\vec v$.
Eq.\eqref{S1} describes the linear response regime corresponding to the so-called Stokes flow or creeping flow, which arises in the low-Reynolds limit.
%\addLL{[comment on the units of $\alpha$ and $u$ being different from the main text]}
We note that in the main text $\alpha$ and $u$ represent scattering rate and thus have the dimension of frequency, % $[\rm{T}]^{-1}$, while
whereas here they describe the rate of change of momentum density through $\p_t p_i+\eta\nabla^2 v_i=\alpha(\vec r) v_i+...$ i.e. they are related with the quantities in the main text as $\alpha\to nm\alpha$, $u\to nmu$.
% have the dimension of frequency times density $[\rm{T}]^{-1}[\rm{L}]^{-3}[\rm{M}]$.

We will start with a single scatterer in a uniform flow. Introducing a stream function to resolve the incompressibility condition, $\vec v=\vec z\times\nabla\psi$, and taking a curl of Eq.\eqref{S1} to eliminate the term $\nabla\phi$, we find that the stream function satisfies
\be\label{eq:psi}
\hat K\psi(\vec r)=0
,\quad
\hat K = \eta(\nabla^2)^2 -\tilde\p_i u(\vec r)\tilde\p_i
\ee
where $\tilde\p_i $  are components of the rotated gradient operator $\tilde\nabla= \vec z\times\nabla$.
We will seek a solution that describes an asymptotically uniform flow perturbed by the scatterer, $\psi(\vec r)=\psi_0(\vec r)+\delta\psi(\vec r)$, where without loss of generality we take the velocity at infinity ${\vec v_0}\parallel \vec x$ and place the scatterer at the origin of the coordinate system. Then $\psi_0(\vec r)=-\vec z\cdot ({\vec v_0}\times \vec r)$. Solving for $\delta\psi$ we find
\be
\delta\psi(\vec r)=\hat K^{-1}\tilde\p_i u(\vec r)\tilde\p_i\psi_0(\vec r)
.
\ee
To elucidate the dependence on viscosity we develop perturbation series in $u(\vec r)$ by writing $K^{-1}=K_0^{-1}+K_0^{-1}VK_0^{-1}+K_0^{-1}VK_0^{-1}VK_0^{-1}+...$ where $K_0=\eta(\nabla^2)^2$, $V=\tilde\p_i u(\vec r)\tilde\p_i$. These series can be analyzed more conveniently in Fourier representation. In particular, it is instructive to compare the first two terms
\be\label{eq:psik}
\delta\psi(k)=\frac{i\tilde {\vec k}_i}{\eta k^4}\lp u_{\vec k} {\vec v_0}_i
-\sum_{\vec q}u_{\vec k-\vec q}  D_{ii'}(\vec q)
u_{\vec q}{\vec v_0}_{i' } +...\rp
\ee
where we defined $D_{ii'}(\vec q)= \frac{\tilde {\vec q}_i\tilde {\vec q}_{i' }}{\eta q^4}$
and used Fourier harmonics $u_{\vec k}=\int e^{-i\vec k\vec r} u(\vec r) d^2r$. Integration over $\vec q$ gives a log divergence at small $\vec q$, a behavior directly related to the well-known Stokes paradox \cite{F,HB}. This divergence has to be  cut off at $q\sim 1/L$, where the length $L$ is set by the distance between scatterers or the system size, whichever is smaller. At large  $\vec q$ the integral is cut off at $q\sim 1/a$, where $a$ is the scatterer size. Replacing $u_{\vec k}$ and $u_{\vec k-\vec q}$ with $u_{\vec k=0}$ and estimating the integral over $\vec q$ as
\be
\sum_{\vec q}D_{ii'}(\vec q)
=\frac1{4\pi\eta}\ln\frac{L}{a}\delta_{ii'}
\ee
we see that the second-order term gives the $\vec k$ dependence identical to that in the first term wherein the value $u_0$ is replaced with $-\frac{u_0^2}{4\pi\eta}\ln\frac{L}{a} $. Extending these observations to higher-order terms we can sum the series and write the result in terms of a renormalized scatterer strength
\be\label{eq:U(eta)}
U(\eta)=\frac{u_0}{1+\frac12u_0\sum_{\vec q}\frac1{\eta \vec q^2}}
\ee
which gives the log dependence
\be\label{eq:mu_U(eta)_new}
%\mu=\frac{N v_F^2}{2n_s n U(\nu)}
%,\quad
%U(\nu)=\frac{U_0}{1+\frac{U_0}{4\pi \nu}\ln \frac{L}{a_*}}
%\mu=\frac{2ne}{n_s U(\eta)}
\quad U(\eta)=\frac{u_0}{1+\frac{u_0}{4\pi\eta}\ln \frac{L}{a}}
\ee
After identifying/rescaling parameters we find that this is identical to the dependence in Eq.\eqref{eq:sigma_log} of the main text,
where $L$ and $a$ are the IR and UV cutoffs discussed above. It is instructive to compare Eq.\eqref{eq:mu_U(eta)_new} with Eqs.(7-7.23) from Ref.\cite{HB}.
%\addLL{[fix values and scales]}

Next we show that the dependence $U(\eta)$ translates into
a suppression of electrical resistance. For that we evaluate the perturbed velocity  $\delta\vec v=\tilde\nabla\delta\psi$ and plug it in Eq.\eqref{S1} to find
the electric field
\be
\vec E(\vec r)=-\nabla\phi(\vec r)=\frac1{ne}\eta\nabla^2\tilde\nabla \delta\psi(\vec r)
\ee
Combining with the result $\delta\psi(k)=\frac{\tilde {\vec k}\cdot{\vec v_0}}{\eta k^4}  U(\eta)$ found above, we obtain
\be\label{eq:E(k)}
\vec E(\vec k)=-\nabla\phi(\vec r)=\tilde{\vec k}\frac{(\tilde {\vec k}\cdot{\vec v_0})}{\vec k^2}  \frac{U(\eta)}{ne}
.
\ee
Notably the electric field depends on $\eta$ only through $U(\eta)$ since $\eta$ and $1/\eta$ in the prefactor cancel out.

The $\vec k$ dependence in Eq.\eqref{eq:E(k)} translates into a $1/r^2$ power law dependence in position space. Importantly, despite the sign-changing angular dependence, the resulting electric field has a nonzero spatial average.
We integrate over $\vec r$ and, assuming an isotropic scatterer, average over the azimuthal angle to obtain a factor $1/2$. This gives
\be
\int \vec E(\vec r) d^2r=\frac{U(\eta)}{2ne}{\vec v_0}
.
\ee
It is straightforward to apply these results to many randomly positioned scatterers.
The average electric field, which is now proportional to  the density $n_s$ of the scatterers, equals
\be
\la\vec E(\vec r)\ra=A^{-1}\int \vec E(\vec r) d^2r= \frac{n_s U(\eta)}{2ne}{\vec v_0}
\ee
where $A$ is system area.
The proportionality relation between the applied electric field and the drift velocity gives an $\eta$-dependent mobility
\be\label{eq:mu_new}
%\mu=\frac{N v_F^2}{2n_s n U(\nu)}
%,\quad
%U(\nu)=\frac{U_0}{1+\frac{U_0}{4\pi \nu}\ln \frac{L}{a_*}}
\mu=\frac{2ne}{n_s U(\eta)}
%\quad U(\eta)=\frac{u_0}{1+\frac{u_0}{4\pi\eta}\ln \frac{L}{a}}
.
\ee
The dependence on the scatterer strength mimics that in Eq.\eqref{eq:sigma_log} of the main text.
% \addLL{[fix values and scales]}
In the viscous limit (small $\eta$ values), the scatterer strength $U(\eta)$ diminishes and the mobility increases, in agreement with the picture discussed above.

% \addLL{Explain how the stream function in Fig.1b was obtained}
From Eq.\eqref{eq:psik} we evaluate the stream function change near each scatterer
\be
    \delta\psi(\vec{x})=U(\eta) {\vec v_0}_i \int \frac{\rm{d}^2 k}{(2\pi)^2} \frac{i\tilde {\vec k}_i}{\eta k^4}e^{i \vec{k}\cdot\vec{x}}
\ee
    The integral has an IR divergence, which we regularize by the finite system size $L$. This gives
\be
\delta\psi(\vec{x})=\frac{U(\eta)}{4\pi\eta} \lp \ln \frac{2L}{|\vec{x}|} \rp \lp \hat{\vec{z}}\cdot(\mathbf{v}_0\times\mathbf{x})\rp
.
\ee
% where $\xi\sim L$ is the IR cutoff. 
Generalizing to many scatterers we find
\be
\psi(\vec x)=\psi_0(\vec x)+\sum_s \delta\psi(\vec x -\vec x_s)
.
\ee
We use the isolines of $\psi(\vec x)$ to plot the streamlines shown in Fig.\ref{fig1}a of the main text.
% The corresponding flow configuration is plotted in Fig.1b.
% [Compare the log enhancement to other log effects, and to $\eta$ dependence from other effects.]

%\addLL{These results from hydrodynamical equation, while giving a very reasonable $\eta$ dependence, do not describe the ballistic-to-viscous crossover. This is so because the starting point is the Stokes equation which is valid in the hydrodynamic limit when all length scales are greater than the electron-electron collision mean free path $\ell_{\rm el-el}$. This includes, in particular, the scatterer radius $a$.( Below we will extend our treatment to Boltzmann equation which captures both the collisionless Knudsen regime and the hydrodynamic regime.)}

Finally, we point out that the log divergences found above, such as the one in Eq.\eqref{eq:psik},  are directly related to the seminal Stokes paradox \cite{F,HB}. The name `Stokes paradox' refers to the simple the there is no non-trivial, steady state solution for the linearized Navier-Stokes equation in a 2D disk geometry, which describes a uniform flow at infinity. 
Physically, this behavior stems from the simple fact that a motion of a body relative to a fluid produces velocity perturbation logarithmically growing with the distance within the framework of the linear equation (\ref{S1}) in 2D. The growth saturates %is stopped 
at the distances where viscous friction is balanced by inertia, which was neglected in  (\ref{S1}). %That leads to the 
For a disk of radius $a$, this yields velocity distribution
\be
v(r)\simeq %\frac{U_\infty}{\ln {\rm Re}^{-1}}
v_\infty \ln(r/a)/\ln ({\rm Re}^{-1})
,
\ee 
where ${\rm Re} =  {v_\infty a}/{\nu}$ is the Reynolds number,  $v_\infty$ is the velocity far away from the disk. % and $a$ is the disk radius.  
Since the friction force is determined by the fluid velocity gradient on the body which is logarithmically small compared to  $v_\infty/a$, 
the drag is logarithmically suppressed \cite{HB,VanDyke,RevVeysey, ProudmanPearson}.
%\begin{equation}
%C_D = -\frac{4\pi}{\rm Re}\frac{1}{\gamma+\frac{1}{2}+\ln{({\rm Re}/4)}}
%\label{drag}
%\end{equation}
%where $\gamma=0.5772...$ is the Euler-Mascheroni constant.
%In deriving the result for $C_D$, the cutoff is introduced in real space and is velocity-dependent, making the solution nonlinear in $U_\infty$.}
%Logarithmic divergences occur in classical fluid dynamics in the resolution of second Stokes paradox, leading to corrections in the drag coefficient $C_D$ of an infinite cylinder (Eq. \ref{drag}, \cite{VanDyke}) which resemble our scatterer and mobility renormalization.
% , which can be put in direct analogy with the resistance caused by a disorder in an electron flow or with the scatterer strength $U(\eta)$ derived in this work.

This behavior is quite similar to that found in our treatment of electron flow in the presence of point-like scatterers. In particular, the logarithmic enhancement of conductivity is a direct analog of the logarithmic drag suppression in the Stokes problem. We note, however, that the two problems differ in one important way. The Stokes paradox is resolved using a velocity-dependent real-space cutoff, generating a logarithmic velocity dependence of the drag coefficient. 
In contrast, our log divergence is cut at the velocity-independent IR and UV scales, giving rise to a velocity-independent mobility. At the same time, parallel to the findings of this work, one can see that reducing the viscosity value $\nu$ enhances the Reynolds number ${\rm Re}$, which reduces the drag coefficient % $C_D$ 
\cite{Tritton}. This is analogous to our scatterer strength $U(\eta)$ down-renormalization.

\end{document}